\documentclass[twocolumn,showpacs,amsmath,amssymb]{revtex4-1}

\usepackage{graphicx}

\usepackage{amsmath,amsbsy}

\newcommand{\be}{\begin{equation}}
\newcommand{\en}{\end{equation}}
\newcommand{\bea}{\begin{eqnarray}}
\newcommand{\ena}{\end{eqnarray}}
\newcommand{\hbo}{\hbox to 1 true cm {\hfill } }

\newcommand*\Bell{\ensuremath{\boldsymbol\ell}}

\begin{document}

\title{Phase diagram of the quantum $O(2)$-model in $2+1$ dimensions }

\author{Kurt Langfeld}
\affiliation{School of Computing \& Mathematics, University of Plymouth,
Plymouth PL4 8AA, UK }

\date{\today}

\begin{abstract}
The quantum $O(2)$ model in $2+1$ dimensions is studied by simulating 
the 3d $O(2)$ model near criticality. Finite densities are introduced
by a non-zero chemical potential $\mu $, and the worm algorithm is used to 
circumvent the sign problem. The renormalisation is discussed in some detail. 
We find that the onset value of the chemical potential coincides with the 
mass gap. The $\mu$-dependence of the density rules out Bose-Einstein 
condensation and might be compatible with an interacting Fermi gas. 
The $\mu $-$T$ phase diagram is explored using the density and the 
magnetic susceptibility. In the cold, but dense regime of the phase 
diagram, we find a superfluid phase.  

\end{abstract}

\pacs{ 11.15.Ha, 12.38.Aw, 12.38.Gc }

\maketitle


\section{Introduction
}
The $O(2)$-model as a statistical field theory in three dimensions has always 
enjoyed great phenomenological importance: to name only a few applications, 
it illuminates the superfluid transition ($\lambda $ -point) of pure
$^4He$~\cite{Ahlers1976},  it describes the electromagnetic penetration depth 
in certain high-$T_c$ superconducting materials~\cite{Kamal1994} and it
might provide insights in the Bose-Einstein condensation of atomic vapour (hard
sphere bosons)~\cite{Gruter1997}. It is well known (see
e.g.~\cite{Gottlob:1993jf}) that the 3d classical theory possesses a second
order phase transition at a critical inverse temperature $\beta _c =
0.45420(2)$ separating the disordered phase at $\beta < \beta_c$ from an 
ordered state for $\beta > \beta_c$. In solid state physics, the atomic
characteristic length $a$ sets the fundamental scale of the theory. 
For later use, we introduce the dimensionless correlation length as the
physical correlation length in units of the atomic scale, i.e.,  
$ \xi _\mathrm{latt} (\beta ) := \xi /a $. It is this correlation length which
diverges at the phase transition. 

\vskip 0.3cm
The $O(2)$-model inherits its name from a global $O(2)$ symmetry, which 
lets us define a (Noether) current. The generalisation of the $O(2)$-model to 
include a finite chemical potential, which favours the proliferation of 
density, has recently attracted a lot of interest: The probabilistic weight
is now complex, and the model inhibits the numerical study with classical
Monte-Carlo methods  based upon importance sampling with respect to a real and
positive probabilistic weight. Thus, the model shares this notorious sign 
problem with the theory of strong interactions, i.e., QCD, at finite baryon
densities. The QCD phase diagram as a function of temperature and chemical
potential is still not known form first principle calculations, and the
prospects are bleak unless a change of paradigm let us address theories with
sign problems on more generic grounds. Such breakthroughs might equally well
be 
made in the context of simplistic theories such as the $O(2)$-model at finite
densities. Over the recent past, quite some progress has been made in this 
direction: Monte-Carlo simulations which are based upon importance sampling
with respect to the density of states, an always real and positive quantity,
might circumvent the sign problem~\cite{Wang2001,Langfeld:2012ah}.  
The so-called complex Langevin approach~\cite{Parisi:1984cs,Karsch1985} 
is based upon a complexification of the fields and might be ideally suited to
address complex action systems. This approach has been largely explored
over the recent years~\cite{Aarts:2008wh,Aarts:2009uq} including the 
$SU(3)$ spin model at finite densities~\cite{Aarts:2011zn}.
Concerns about its reliability have been raised
recently, since the approach, although it converges, does not give the
correct answer in certain cases. While the Bose gas is a spectacular success
story for the complex Langevin approach~\cite{Aarts:2008wh,Aarts:2009hn}, the 
approach fails to produce the correct answer for a close relative, 
the $O(2)$-model in certain regions of parameter space~\cite{Aarts:2010aq}. 
Progress has been made recently by devising criteria for correctness of the
method~\cite{Aarts:2011sf,Aarts:2011ax}. Another promising idea, firstly put
forward by  
Chadrasekharan in~\cite{Chandrasekharan:2008gp}, is based upon a
reformulation of the theory in terms of new, potentially non-local variables
in a sign-problem free manner. In particular, the so-called world-line or worm 
algorithm~\cite{Prokof'ev:2001zz,Prokof'ev:2009xw} 
has been identified as an approach with a wide range of applications including
theories with sign problems. As pointed out by Gattringer, the so-called
$SU(3)$ spin model, which is motivated from dense QCD in the limit of heavy
quarks, is accessible by means of worm (also called flux type)
algorithms even at finite
densities~\cite{Mercado:2011ua,Gattringer:2011gq,Mercado:2012ue,Gattringer:2012jt}.  
Most relevant for the study here is the flux representation of the 
$O(2)$-model~\cite{PhysRevE.67.015701,PhysRevE.68.026702}. A thorough 
analysis of the classical model at finite densities 
has been presented by Chadrasekharan in~\cite{Banerjee:2010kc}. 
Finally, the so-called fermion bag approach has recently been put forward 
in~\cite{Chandrasekharan:2009wc,Chandrasekharan:2011mn} and might have the
potential to gain new insights into fermionic theories with sign problems. 

\vskip 0.3cm 
Although a range of classical theories with sign-problems have been explored
in the recent past, to the best of our knowledge such an investigation of 
a finite density quantum field theory has not yet been performed. 
In the present paper, we will bridge this gap and systematically explore the
quantum limit of the $O(2)$-model at finite densities. 
For this purpose, we rely to a large extent on the flux type
algorithm~\cite{Banerjee:2010kc} which will give us access  
to the full phase diagram as a function of temperature and chemical
potential. Generic quantum phenomena such as superfluidity will be described 
by a first principle simulation.

\vskip 0.3cm
On a practical note, the quantum limit of the model is accessible 
using simulations of the classical theory and by performing a detailed 
scaling analysis near the critical point. This will then reveal the properties 
of the quantum $O(2)$ theory in $2+1$ dimensions. Let us explore the
correspondence between classical and quantum theory a bit further: 
In the latter context, the inverse temperature  $\beta $ is
re-interpreted as the spin-spin coupling strength, 
and the temperature $T$ is now attained by the extent of the torus in ``time''
direction. The inverse correlation length is now interpreted as the mass gap,
$m = 1/\xi$. This mass takes over the role of the fundamental scale of the 
quantum field theory and is kept fixed under a change of the coupling strength 
$\beta $. The formerly introduced atomic length $a$ (also called the 
lattice spacing in the quantum context) merely plays the role of a regulator. 
This regulator $a$ now necessarily depends on the coupling strength: 
$ m \, a(\beta ) = 1/\xi _\mathrm{latt} (\beta ) $. Tuning the coupling 
$\beta $ to the critical value (at a fixed value for $m$) implies that the
lattice spacing tends to zero. This then installs the continuum limit 
of the quantum theory.

\section{Model set-up and symmetries}

In $2+1$ dimensions, the $O(2)$-model, also called the $xy$-model, is
formulated on a cubic grid of size $V=N_t N_x N_y$. Dynamical degrees of
freedom are 
the angles $\phi_x$ which are associated with the sites $x$ of the lattice and
which parametrise the unit vectors
\be
\vec{n}_x = \Bigl( \cos(\phi _x), \sin (\phi _x) \Bigr)^T \; .
\label{eq:1}
\en
Action and partition function are given by
\bea
S_0 &=& \sum _{\ell =\langle x, \nu \rangle} \vec{n}_x^T \vec{n}_{x+\nu}
\; = \; \sum _{\ell =\langle x, \nu \rangle} \cos(\phi_x- \phi_{x+\nu}) \; ,
\label{eq:2} \\
Z &=& \int {\cal D}\phi_x \; \mathrm{exp} \Bigl\{ \beta \, S_0[\phi] \Bigr\} ,
\label{eq:3}
\ena
where $\beta $ is the coupling constant, and
where $x+\nu $ is the neighbouring site of $x$ in $\nu $ direction.
The sum in (\ref{eq:2}) extends over all links $\ell $ of the lattice. 
The model possesses a global $O(2)$ symmetry,
\be
\vec{n}_x \to O(\phi) \vec{n}_x \; \Rightarrow \; \phi _x \to \phi_x + \phi \; ,
\label{eq:4} 
\en
where $O(\phi) \in O(2)$. 
This displacement symmetry gives rise to the Noether current:
\be
j_\mu(x) \; \propto \; \sin ( \phi_x - \phi _{x+\mu} ) \; ,
\label{eq:5}
\en
Indeed, defining the divergence of the current by
\be
\Delta _\mu \, j_\mu (x) \; \propto \;
\sum _\mu \Bigl[ j_\mu (x) - j_\mu (x-\mu) \Bigr] ,
\label{eq:6}
\en
and using the identity (for any $x_0$)
$$
\int {\cal D}\phi_x \; \frac{\partial }{\partial \phi _{x_0} } \;
\mathrm{exp} \Bigl\{ \beta \, S_0[\phi] \Bigr\} \; = \; 0 \; ,
$$
we find that any number of insertions of the divergence (\ref{eq:6}) into the
partition function causes the integral to vanish, e.g., 
\be
\int {\cal D}\phi_x \; \Delta _\mu \, j_\mu (x_0) \;
\mathrm{exp} \Bigl\{ \beta \, S_0[\phi] \Bigr\} \; = \; 0 \; .
\label{eq:7}
\en
The model can then be readily generalised to finite charge densities, by
introducing a chemical potential $\mu $ to the action in the
usual way~\cite{Hasenfratz:1983ba}:
\be
S [\phi ]
\; = \; \sum _{\ell = \langle x, \nu \rangle } \cos(\phi_x- \phi_{x+\nu}
- i \, \mu \, \delta _{\nu 0 } ) \; .
\label{eq:8}
\en
Thus, the imaginary part of the action is proportional to the
conserved charge $j_0(x)$ (\ref{eq:5}), i.e.,
\bea
S [\phi ] &=& \sum _{\ell = \langle x, \nu \rangle }
\Bigr[ \mathrm{cosh}(\mu) \, \cos(\phi_x- \phi_{x+\nu})
\label{eq:9} \\
&-& i \, \mathrm{sinh}(\mu)  \; \delta _{\nu 0 }  \;
\sin(\phi_x- \phi_{x+\nu}) \Bigr] \; .
\nonumber
\ena
For large values of the coupling $\beta $, the global $O(2)$ symmetry
(\ref{eq:4}) is spontaneously broken in the infinite volume limit.
The operator $\cos (\phi_x)$ is not invariant under the symmetry
transformation, and its expectation value serves as an order parameter
in the infinite volume limit. Note, however, that for any finite
volume $V$, the expectation value 
\be
\Bigl\langle \cos (\phi _x) \Bigr\rangle \; = \;  0 ,
\label{eq:11}
\en
i.e., it vanishes for any value of the coupling $\beta $. This 
can be easily seen by performing the variable transformation 
\be 
\phi _x \to \phi _x + \varphi 
\label{eq:11a}
\en 
in the integral (\ref{eq:3}). 
In order to trace out any indications for spontaneous symmetry breaking
at finite volumes, we add a source term to the action which breaks the
$O(2)$ symmetry explicitly. We then study the response of the
expectation value in (\ref{eq:11}) to variations of the external source.
Here, we choose
\be
Z [j] \; = \;  \int {\cal D}\phi_x \; \mathrm{exp} \Bigl\{ \beta \, S[\phi]
\, + \, j \, \sum _x \cos (\phi_x) \; \Bigr\} .
\label{eq:10}
\en
The response function is defined by
\be
R_j(\beta) \; = \; \Bigl\langle \cos (\phi _x) \Bigr\rangle \; = \;
\frac{1}{V} \, \frac{ \partial } {\partial j} \, \ln Z[j] \; .
\label{eq:11b}
\en
To decide whether the $O(2)$ symmetry is spontaneously broken, we take
the limits and find
$$
\lim _{j \to 0 } \lim _{V \to \infty } R_j(\beta ) \;  \left\{
\begin{array}{ll}
= \; 0 & \hbox{Wigner-Weyl realisation } \\
\not=0 & \hbox{spontaneous sym.~breaking  , }
\end{array} \right.
$$
where the order of the limits is crucial.
For this program, it is sufficient to study small perturbations $j$ only,
which leads us in leading order and finite volumes to
\bea
R_ j &=& \chi (\beta ) \, j \; + \; {\cal O}(j^2) \; ,
\label{eq:16} \\
\chi (\beta) &=& \frac{1}{V} \frac{ \partial^2 }
{\partial j^2} \ln Z[j] \vert_{j=0}
=  \Bigl\langle \cos (\phi _{x_0}) \sum _x \cos (\phi _x) \Bigr\rangle ,
\nonumber 
\ena
where $\chi $ is (half of) the so-called magnetic susceptibility.
If $\chi $ remains finite in the infinite volume limit, the
response function vanishes in the limit of vanishing source $j$, and
the system is in the Wigner-Weyl phase. Hence, a diverging magnetic
susceptibility in the infinite volume limit is a necessary condition for
the spontaneous symmetry breakdown.

\begin{figure}
\includegraphics[width=8cm]{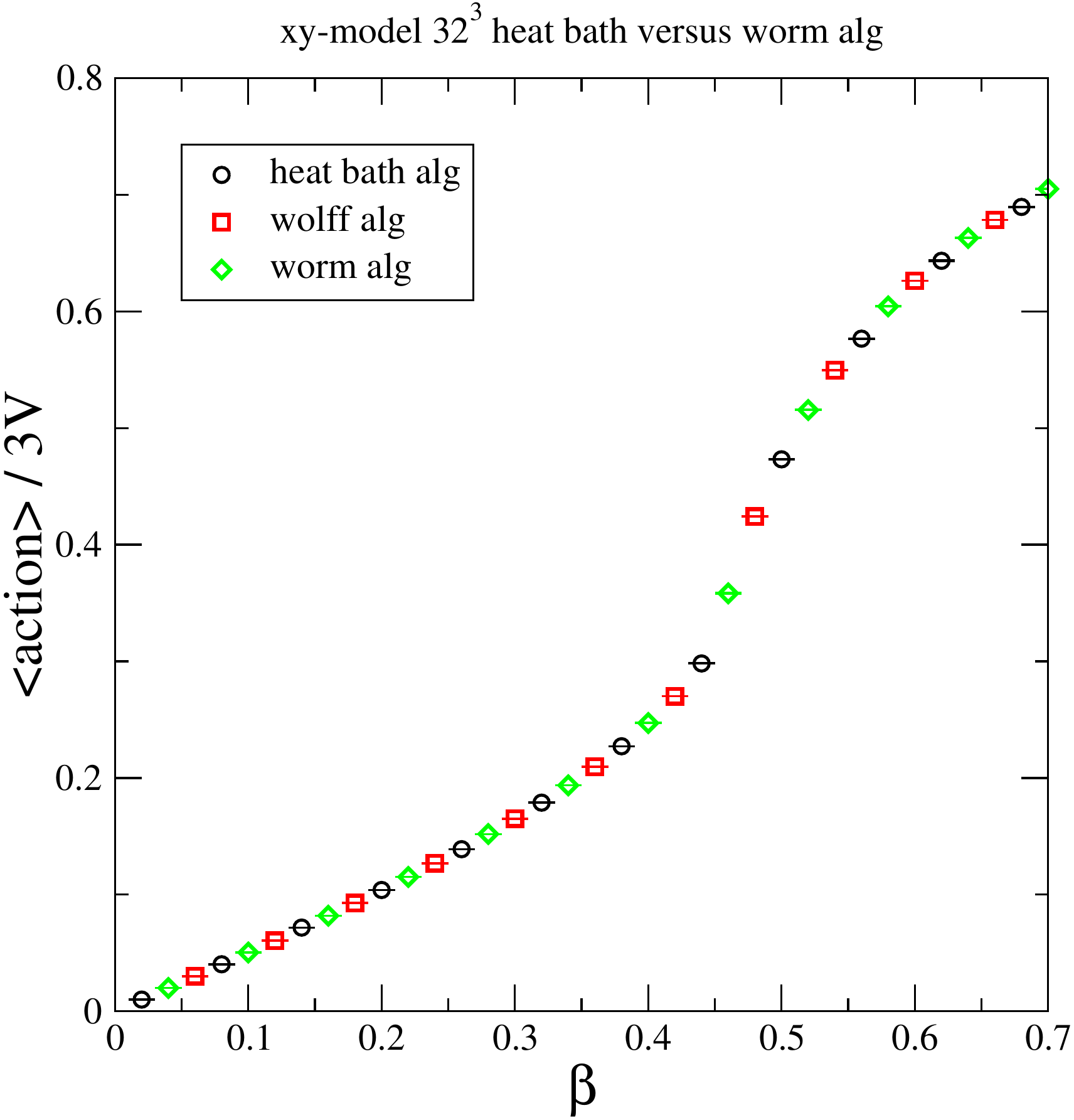}
\caption{\label{fig:1} The average action as a function of $\beta $ for the 3d
  $O(2)$   model using a $32^3$ lattice. }
\end{figure}
Note also that $\chi $ can be viewed as the expectation value for
$\sum \cos (\phi)$ where the factor $ \cos (\phi _{x_0})$ acts as a
reference against which the spin orientation is counted. 
The magnetic susceptibility can be written in a manifestly $O(2)$ invariant 
form. To this end, we rewrite the 
expectation value on the right hand side of (\ref{eq:16}) as
$$
\frac{1}{2} \sum _x \Bigl\{ \;
\Bigl\langle \cos (\phi_x - \phi _{x_0})  \Bigr\rangle +
\Bigl\langle \cos (\phi_x + \phi _{x_0})  \Bigr\rangle \; \Bigr\} . 
$$
Using the variable transform (\ref{eq:11a}), we conclude that 
the latter expectation value vanishes for $j=0$ (at any
finite volume). Thus, we can write the magnetic susceptibility as
\be
\chi (\beta) \; = \; \frac{1}{2} \sum _x \Bigl\langle
\cos (\phi_x - \phi _{x_0})  \Bigr\rangle \; ,
\label{eq:17}
\en
which is manifest $O(2)$ invariant. Introducing the spin correlation function,
by 
\be
G(x-y) \; = \; \Bigl\langle \vec{n}_x \cdot \vec{n}_y \, \Bigr\rangle \; ,
\label{eq:18}
\en
the magnetic susceptibility can be easily related to the integrated
spin correlation function:
\be
\chi (\beta) \; = \; \frac{1}{2} \, \sum_x \, G(x-x_0) \; .
\label{eq:19}
\en
The latter identity explains the role of $\chi (\beta) $ as an order
parameter for the spontaneous breaking of the $O(2)$ symmetry:
in the disordered phase, the spin correlation function exponentially decreases
over the distance of the correlation length $\xi $, and
$\chi (\beta) $ is independent of the system size. If the correlation
length near criticality exceeds the system size, we find by means of
the sum at the right hand side of (\ref{eq:19}) that
$\chi (\beta) $ diverges with the volume.

\section{The statistical field theory }

\subsection{Algorithms and first results }

\subsubsection{Wolff cluster algorithm}
\begin{figure}
\includegraphics[width=8cm]{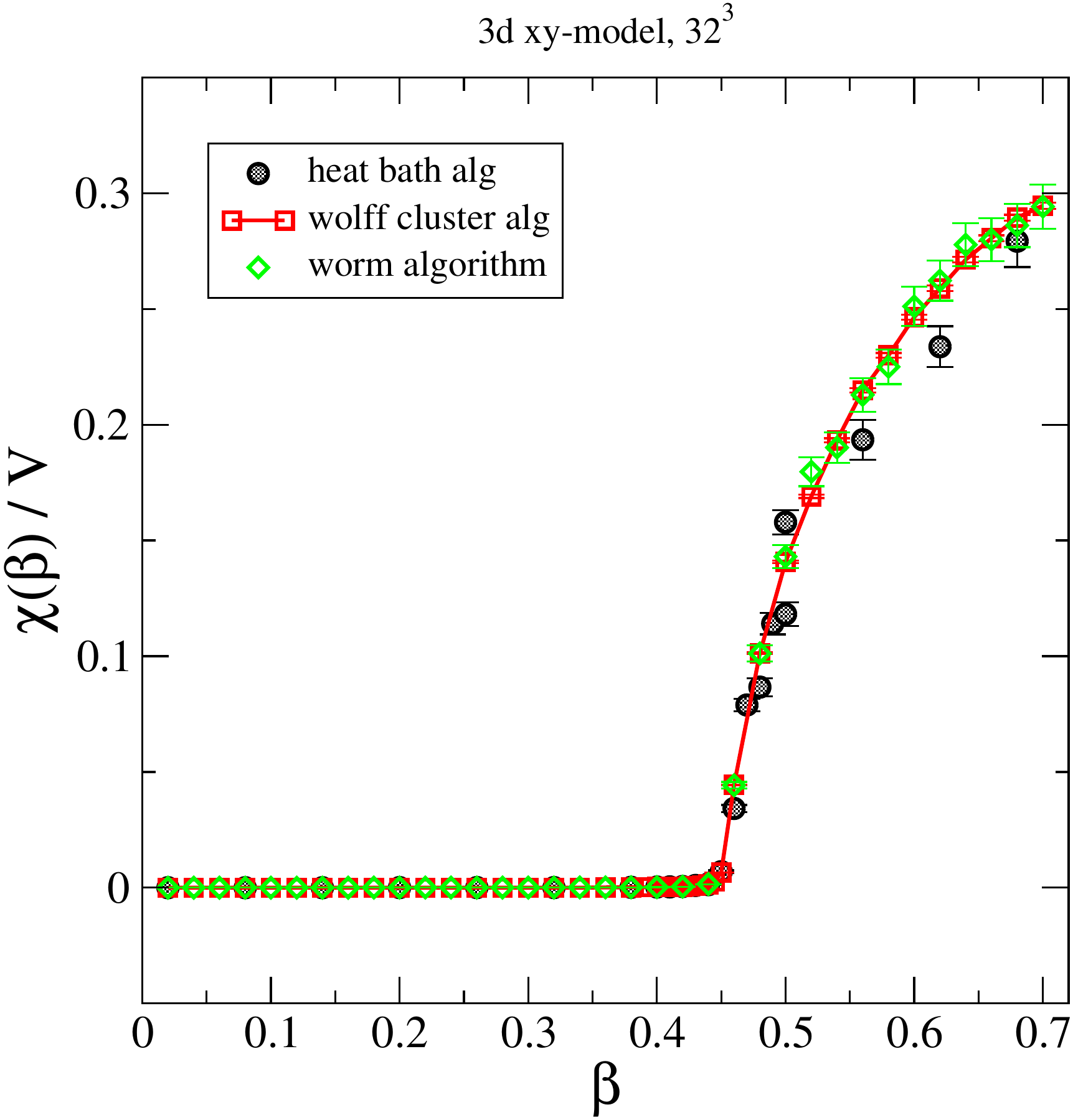}
\caption{\label{fig:2} The magnetic susceptibility per volume as a function of
  $\beta $   for the 3d   $O(2)$   model using a $32^3$ lattice. }
\end{figure}
\begin{figure}
\includegraphics[width=8cm]{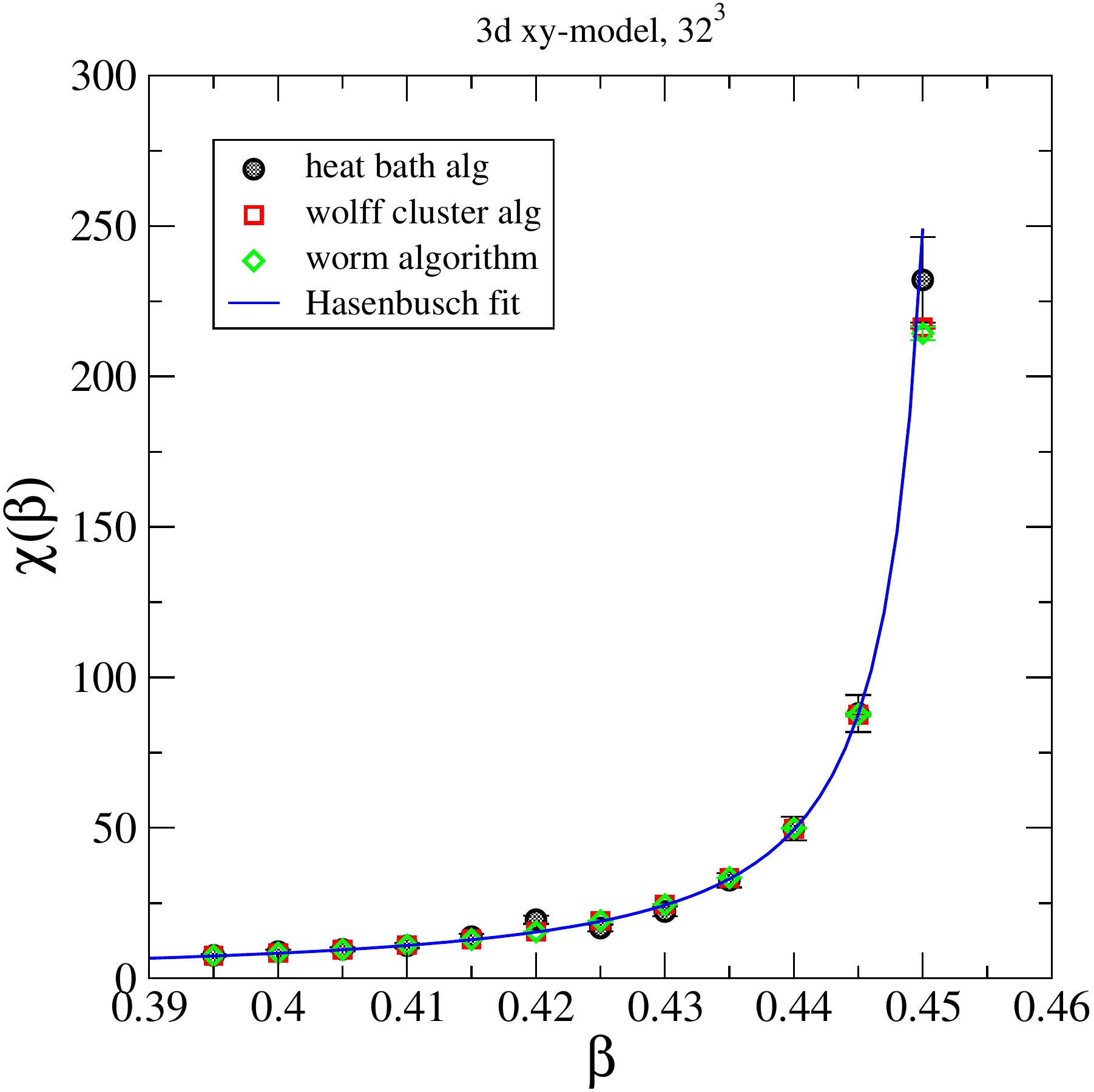}
\caption{\label{fig:3} Detail of the magnetic susceptibility as a function of
  $\beta $ close but smaller than the critical value ($32^3$ lattice). }
\end{figure}
For vanishing chemical potential, i.e., $\mu=0$, and vanishing source $j$,
the  Wolff cluster algorithm~\cite{1989NuPhB.322..759W}
offers a very efficient way to simulate the $O(2)$-model. The reason is 
that the clusters, which are updated in one Wolff step, grasp the physics of
the model. This implies that the auto-correlation time depends only weakly 
on the system size as reflected by a dynamical critical exponent close to zero.
The algorithms proceeds as follows:
\begin{itemize}
\item[1.] Choose a random unit vector $\vec{r}$;
\item[2.] Activate the link between two neighbouring sites $x$ and $y$ with
probability
$$
P(x,y) = 1 - \exp \left\{ \mathrm{min}\left[0, \,  - 2 \, \beta \,
    (\vec{n_x} \cdot \vec{r}) \,  (\vec{n_y} \cdot \vec{r}) \, \right]
\right\} ;
$$
\item[3.] Select a random site $x_0$ and find the subset $C$ of all spins which
are connected to $\phi _{x_0}$ by activated links
\item[4.] Replace
$$
\vec{n} _x \to \vec{n} _x \, - \, 2 \, (\vec{n_x} \cdot \vec{r}) \, \vec{r}
\hbo \forall \; x \in C \; .
$$
\end{itemize}
Observables are best calculated using the so-called improved 
estimators~\cite{Hasenbusch:1989kx,Wolff:1989hv}. In particular, the 
magnetic susceptibility can be directly related to the cluster 
properties and as such calculated ``on the fly'': 
\be 
\chi (\beta ) \; = \; \Bigl\langle \frac{ 1 }{V_c}  
\Bigl( \sum _{x\in C} \vec{r} \cdot \vec{n}_x \Bigr)^2 \, \Bigr\rangle \; , 
\label{eq:19a} 
\en 
where $C$ denotes a particular Wolff-cluster, and $V_c$ is the number 
of sites belonging to this cluster $C$. 

\subsubsection{1-shot heat bath algorithm}

Heat bath algorithms offer an easy access to expectation values formulated 
in the terms of the original spin degrees of freedom. We here use the 
heat bath algorithm to test and benchmark the flux algorithm outlined in 
the next subsection. 
For this purpose, we have developed a heat bath algorithm with 100\%
acceptance rate, which has been inspired by the method from Bazavov and
Berg in~\cite{Bazavov:2005zy}. In step 1, we choose the spin $\phi _{x}$ for
the local update step. If $\langle xy \rangle $ denotes the links joining
the sites $x$ and $y$, the action can be written as
\bea
\Bigl[ && \beta \, \sum _{y\in \langle xy \rangle}  \cos (\phi _y) \, + \, j
\Bigr] \, \cos (\phi _x) \; + \;
\nonumber \\
&& \Bigl[ \beta \, \sum _{y\in \langle xy \rangle}  \sin (\phi _y) \Bigr]
\, \sin (\phi _x) \; = \; A \, \cos(\phi _x - \psi ) ,
\nonumber
\ena
where $A$ and $\psi $ depend on $\beta $, $j$ and the neighbouring spins.
In step 2, we define
$$
f(\phi_x ) \; = \; \int _{-\pi} ^{\phi_x} \exp \Bigl\{ A \, \cos(v - \psi )
\Bigr\} \; dv \; .
$$
We then choose a random number $u \in [0,1]$ and solve the equation
$$
f(\pi) \; u \; = \; f(\phi_x )
$$
for the new value $\phi _x$ for the spin at site $x$.
In practice, the integration was performed using the Gauss Legendre
method, while the non-linear equation was solved using bisection.

\subsubsection{Worm algorithm \label{sec:worm}} 

For finite values of the chemical potential, i.e., $\mu \not= 0$, the action
(\ref{eq:8}) possesses imaginary parts. The theory is hampered by the infamous
sign problem and cannot be simulated at reasonably sized lattices using
the cluster or the heat bath algorithm. The $O(2)$-model is one of the rare
cases where simulations are nevertheless feasible by using the so-called
worm
  algorithm~\cite{Prokof'ev:2001zz,Syljusen:2002zz,Banerjee:2010kc,Prokof'ev:2009xw}. 
The algorithm has been thoroughly discussed in~\cite{Banerjee:2010kc}. 
We here briefly review the algorithm for the $O(2)$-model at finite chemical
potential before we take it a step further to derive the Kramers-Wannier
duality in the next section. Using the identity
\be
\mathrm{e} ^{ \beta \, \cos(\phi ) } = \sum _{k=-\infty}^\infty
I_k(\beta ) \; \mathrm{e}^{i k \phi } \; ,
\label{eq:20}
\en
where $I_k=I_{-k}$ is the modified Bessel function of the first kind,
the partition function (\ref{eq:10}) becomes
\bea
Z[0] &=& \int {\cal D}\phi _x \; \mathrm{e} ^{ \beta \, S[\phi]} = \int
{\cal D} \phi _x \;
\nonumber \\
&& \sum _{\{k_\ell \}} \prod_{\ell=\langle xy\rangle } I_{k_{\ell}}(\beta ) \;
\mathrm{e}^{\mu \delta _{\nu 0} \, k_\ell } \;
\mathrm{e}^{i k_\ell (\phi_x -\phi_y ) }
\nonumber \\
&=& \sum _{\{k_\ell \}} \prod_{\ell } I_{k_{\ell}}(\beta ) \;
\mathrm{e}^{\mu \delta _{\nu 0} \, k_\ell } \; \prod _x \; \delta\left(
\sum_{\ell \in x} k_\ell \right) \; ,
\label{eq:21}
\ena
where $\nu $ specifies the direction of the link $\ell $. 
The notation $\ell \in x$ refers to all links which are attached
to the site $x$:
$$ 
\sum_{\ell \in x} k_\ell \; = \; \sum_\nu \Bigl( k_{x,\nu} - 
k_{x-\nu,\nu} \Bigr). 
$$
If an integer ``flux'' $k_\ell$ is associated with each
link $\ell $ of the lattice, the $\delta $-function constraint in
(\ref{eq:21}) implies that the total (directed) flux which enters a site must
vanish.

Including the case of a finite chemical potential $\mu $, the worm
algorithm has been widely used in~\cite{Banerjee:2010kc} to explore
properties of the $O(2)$-model. The algorithm works as
follows~\cite{Banerjee:2010kc}:
\begin{itemize}
\item[1.] Pick a starting point $x$ and set the counter $c=0$
\item[2.] Choose at random one of the 6 directions (including negative ones);
let $y= x + \nu $ be the neighbouring site associated with the chosen
direction ($\nu = \pm 1, \pm 2, \pm3 $) and $\ell $ the associated link
$\langle xy \rangle =\ell $.
\item[3.] If a positive direction was selected, with
probability $ \exp ( \mu \delta _{\nu 0}) I_{k_\ell +1} (\beta ) /
I_{k_\ell} (\beta ) $ change $k_\ell $ to
$k_\ell + 1$ and move to the neighbouring site $y$.
Otherwise stay. In any case, increase $c \to c+1$.

If a negative direction was selected, with
probability $ \exp ( -\mu \delta _{\nu 0}) I_{k_\ell -1} (\beta ) /
I_{k_\ell} (\beta ) $ change $k_\ell $ to
$k_\ell - 1$ and move to the neighbouring site $y$. 
Otherwise stay. In any case, increase $c \to c+1$.

\item[4.] Stop if the current position coincides with the starting position,
i.e., if $y=x_s$.

\end{itemize}

\begin{figure}
\includegraphics[width=8cm]{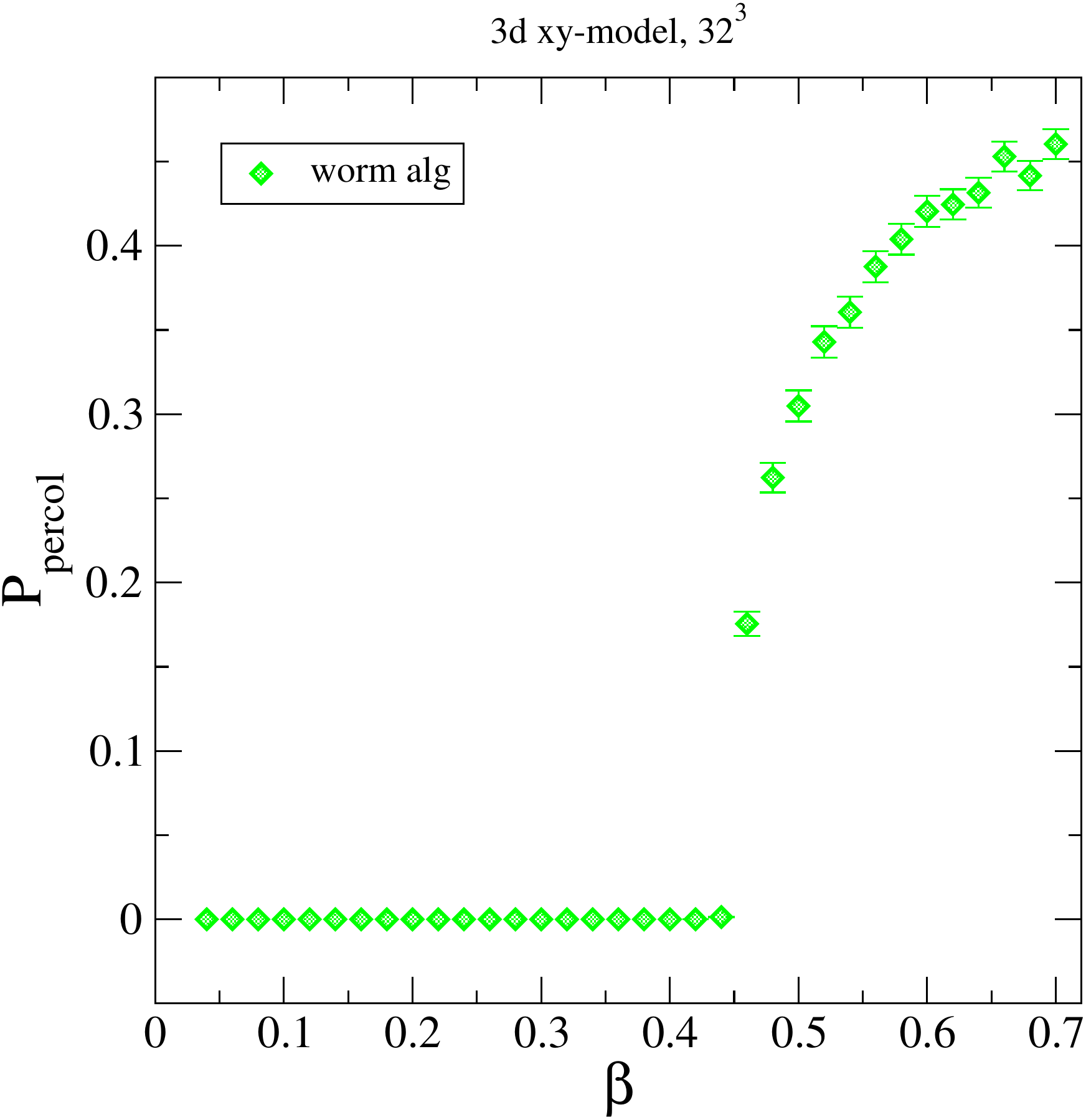}
\caption{\label{fig:4} The percolation probability $P_\mathrm{percol} $
  (\ref{eq:27}) as a function of  $\beta $ ($32^3$ lattice).}  
\end{figure}
The average action is easily obtained by 
\bea
A(\beta ) &=& \frac{\partial }{\partial \beta } \, \ln Z 
\; = \; \sum _{\ell } \, \left\langle 
\frac{ I_{k_\ell }^\prime (\beta ) }{I_{k_\ell} (\beta ) } \right\rangle \; . 
\label{eq:23} 
\ena
It has been argued in~\cite{Banerjee:2010kc} that the magnetic susceptibility 
is given in terms of the average number of counts $c$: 
\be 
\chi (\beta ) \; = \; \frac{1}{2} \, \langle c \rangle \; . 
\label{eq:24} 
\en

\subsubsection{Action average and magnetic susceptibility } 

High precision simulations of the model have been reported in the
literature for zero chemical potential 
e.g.~in~\cite{Hasenbusch:1989kx,Campostrini:2000iw} and for finite densities 
in~\cite{Banerjee:2010kc}. We will benchmark the outcome of our algorithms 
against these findings before we will move on to study the quantum field 
theory (QFT) limit at finite densities in the next section. 
We will employ 
\begin{itemize}
\item the Wolff algorithm to obtain the correlation length in units of 
the lattice spacing, which will set the QFT scale, 
\item the worm algorithm for the simulations at finite densities, 
\item the heat bath algorithm to crosscheck the numerical results. 
\end{itemize} 
To begin with, we have compared all three algorithms by calculating the average
action as a function of the coupling  $\beta $ using a $32^3$ lattice. 
Our findings are shown in figure~\ref{fig:1}. 
We find a nice agreement for all three algorithms for a range of $\beta $
values covering the disordered phase for $\beta < \beta _c$, 
$\beta _c = 0.45420(2)$~\cite{Hasenbusch:1989kx}, as  well as 
the ordered phase $\beta >\beta _c$. Our results for the magnetic 
susceptibility $\chi $ are shown in figures \ref{fig:2} and \ref{fig:3}. 
The transition from the Wigner-Weyl phase to the phase with a spontaneously 
broken $O(2)$ symmetry is clearly visible in figure~\ref{fig:2}. 
As expected the Wolff and the worm algorithm work flawlessly in the broken
phase while the results from the heat bath algorithm are plagued by large 
autocorrelation times. The region of $\beta $ just a bit smaller than 
$\beta _c$ is of particular interest for the QFT limit. Here, all three 
algorithms perform well (see figure~\ref{fig:3}). A good agreement 
with the fit 
\bea 
\chi (\beta ) &=& \chi _0 \, \Bigl( 1 - \frac{\beta }{\beta _c} 
\Bigr)^{-\gamma }, \hbo \beta \stackrel{<}{_\sim } \beta _c \; , 
\label{eq:25} \\
\chi _0 &=& 0.5045(1) , \; \; \; 
\beta _c = 0.45420(2)  , \; \; \; \gamma = 1.324(1) \; , 
\nonumber 
\ena 
presented in~\cite{Gottlob:1993jf,Hasenbusch:1989kx} is observed 
(note there is a factor of $1/2$ difference in the definition of our magnetic 
susceptibility and that in~\cite{Gottlob:1993jf,Hasenbusch:1989kx}).

\subsection{Winding sectors and dual theory } 

\subsubsection{Flux percolation} 
\begin{figure}
\includegraphics[width=8cm]{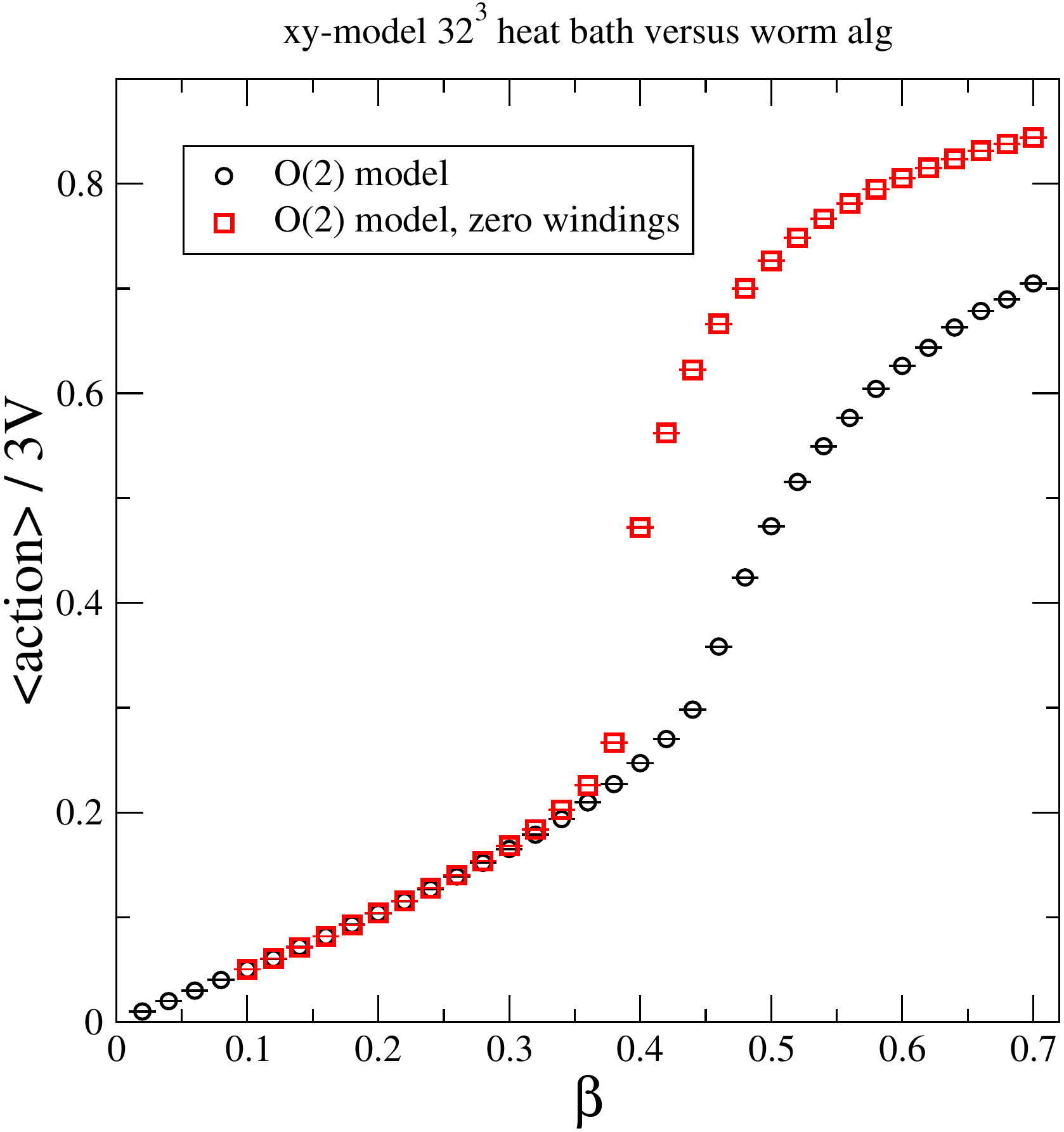}
\caption{\label{fig:5} action density for the standard 3d-O(2) model in
  comparison with model with winding number $n_w=0$ only. }  
\end{figure}
By virtue of the periodic boundary conditions, the closed flux lines of the 
worm algorithm might wind around the torus. Since we currently work with 
a symmetric lattice of equal size $N$ in all directions, we confine ourselves to
the study the winding in time direction. The winding number $n_\mathrm{w} $ can
be calculated during the construction of a particular flux line. Here, we only
need to keep track how often the flux line leaves the torus at the $N_t=N$
and  $N_t=1$, respectively. Every time the flux line leaves the torus at
$N_t=N$  in positive time direction, we increase $n_\mathrm{w} $ by one unit 
while if the flux line leaves the torus at $N_t=1$ negative time direction, we 
decrease $n_\mathrm{w} $ by one unit. One naively might expect that the
average winding number is conclusive on the winding of the flux lines. 
This is, however, not true due to the abundance of small size flux lines 
which all carry $n_\mathrm{w}=0 $. The more interesting quantity is the 
percolation probability $P_\mathrm{percol}$: if we choose randomly a 
particular link of a flux line configuration, how big is the probability that
this link is a part of a flux line which at least winds once through the
torus? If $L$ denotes the number of links of a particular flux line and if 
\be 
L_w \; = \; \left\{ \begin{array}{ll}
L & \hbox{for} \; n_w \not= 0 \\
0 & \hbox{else} 
\end{array} \right. \; , 
\label{eq:26} 
\en 
the percolation probability is given by 
\be 
P_\mathrm{percol} \; = \; \langle L_w \rangle \; / \;  \langle L \rangle \; .
\label{eq:27} 
\en 
Figure~\ref{fig:4} shows our numerical result as a function of $\beta $. 
Note that due to the integration of the $\phi _x$ in (\ref{eq:21}) all 
reference to the global $O(2)$ symmetry has been lost in the flux line 
representation of the partition function. Here, we find empirically that 
the flux lines do not percolate in the Wigner-Weyl phase while the flux 
lines start winding around the torus in the phase with spontaneously broken
symmetry. This is intuitively clear by noting that the correlation length 
diverges for $\beta \to \beta _c$. In the flux representation, a finite
average cluster size implies a finite correlation length by virtue of
(\ref{eq:36}). Hence, the cluster size necessarily diverges at the critical
coupling implying percolation in the finite volume system.

\subsubsection{Kramers-Wannier duality }

The reformulation of a classical field theory in terms of flux variables 
has been dubbed the {\it dual formulation} in the literature (see 
e.g.~\cite{Schmidt:2012uy}). By its nature, this reformulation is 
non-local, using the original lattice, and it is therefore quite different from 
the generic Kramers-Wannier duality, which employs local degrees of freedom 
related to the dual lattice. Starting from the worm algorithm as formulated 
in subsection~\ref{sec:worm}, we here derive the dual formulation of 
the dense $O(2)$ model in the Kramers-Wannier sense. We will find 
a close relation to the flux formulation: the dual theory turns out to be 
a gauge theory with the fluxes being the non-local but gauge invariant 
degrees of freedom of this formulation. 

\vskip 0.3cm 
Starting point is the flux formulation (\ref{eq:21}). Our aim 
will be to resolve the $\delta $-function constraint on the flux elements 
$k_\ell $ using local variables. Naturally, the emerging dual theory is free
of the sign problem at finite chemical potential as the flux line
formulation (\ref{eq:21}) is. To this end, each element of the lattice, i.e.,
site $x$, link $\ell $, plaquette $p$ and elementary cube $c$, is mapped in
the usual way to the dual lattice: 
$$ 
x \to \mathbf{c} , \hbo \ell \to \mathbf{p}, \hbo p \to \Bell , \hbo
c \to \mathbf{x} \; , 
$$
where the ``bold'' quantities denote the entities on the dual lattice. 
Fields depending on the entities are mapped as well, e.g., 
$$ 
k_\ell \to k_{\mathbf{p}} \; . 
$$
The $\delta $-function constraint in (\ref{eq:21}) then becomes on the 
dual lattice: 
\be 
\sum _{\ell \in x } k_\ell =0 \; \; \to \; \; 
\sum _{\mathbf{p} \in \mathbf{c} } k_{\mathbf{p}} = 0 \; , 
\label{eq:30}
\en
where $\mathbf{p} \in \mathbf{c}$ addresses all plaquettes $\mathbf{p} $ 
which build up the faces of the cube $\mathbf{c}$. The crucial point is that
this constraint can be solved {\it locally} for the theory with zero
windings. $ k_{\mathbf{p}}$ are plaquette variables on the dual lattice, and 
the latter constraint (\ref{eq:30}) is reminiscent of the Bianchi identity of 
a $Z$-gauge theory. Hence, we solve the constraint by introducing the 
integer valued link field $Z _{\Bell} $ of the dual lattice by 
\be 
k_{\mathbf{p}} = 
\sum _{\Bell \in \mathbf{p} } Z _{\Bell} 
\label{eq:31}
\en
The definition of the links $Z _{\Bell }$ which represent 
the plaquette $k_{\mathbf{p}}$ is not unique. The gauge transformed links, 
$$ 
 Z ^\Omega _{\Bell=\mathbf{x\mu}} \; = \; \Omega (\mathbf{x}) \, + \, 
 Z _{\Bell=\mathbf{x\mu}} \, - \,  \Omega (\mathbf{x+\mu}) \; , 
$$
where $\Omega (\mathbf{x})$ is integer valued, still give rise to 
the same plaquette in (\ref{eq:31}). A $Z$-gauge symmetry arises in the link
representation of the dual theory. The partition function (\ref{eq:21}) of
the dual theory can be written as:
\bea
Z[0] &=& \sum _{\{Z_{\Bell} \}} \; 
\prod_{\mathbf{p} } I_{k_{\mathbf{p}}} \; 
\exp \Bigl\{ \mu \, k_\mathbf{p^\prime}
\Bigr\} \; , 
\label{eq:32}
\ena
where $k_\mathbf{p^\prime} = k_\mathbf{p} $ for spatial plaquettes and 
$k_\mathbf{p^\prime}=0$ for time-like plaquettes. 
The dual formulation sheds light onto the ``worms'' of the flux line 
representation. By virtue of (\ref{eq:31}), this flux is given by the
plaquettes of the dual $Z$-gauge theory. On the other hand, the dual plaquette
defines the vorticity of the dual theory. Vorticity is conserved 
by the Bianchi identity, and non-trivial vorticity comes form closed 
staples of plaquettes. Hence, we find that the closed flux lines, i.e., the
``worms'', appear to be the vortices of the dual formulation. 

\vskip 0.3cm 
So far, we have established a {\it local} equivalence between the 
standard $O(2)$ model and the $Z$-gauge theory on the dual lattice. 
For a complete correspondence, we also have to match the boundary conditions 
of both formulations. This a non-trivial task: let us consider, e.g., 
periodic boundary conditions for the dual theory. 
If $\cal A$ is the maximal planar surface in the $xy$-plane and $({\cal A})$
its boundary, we find that the total flux through this plane 
necessarily vanishes: 
$$ 
\prod _{\Bell \in ( {\cal A} ) } Z_{\Bell} \; = \; 0 \; . 
$$ 
On the other hand, the standard formulation of the $O(2)$-model with 
periodic boundary conditions does {\it not} restrict the worm 
configurations to the sector of zero winding in $z$-direction. 
Obviously, we have to give up periodic boundary conditions for the 
$Z$-gauge theory if we want correspondence. If $n_w$ is the worm winding
number, the corresponding $Z$-gauge theory simulating this sector features 
non-trivial boundary conditions such as 
\bea 
Z_2(x=1,1,1) &=& Z_2(x=N_x,1,1) + n_w, 
\nonumber \\
\hbox{ (periodic else),} && 
\nonumber 
\ena 
where the subscript ``$2$'' indicates the y-direction of the link. 
The issue, however, is that it is a priori unknown with which weight 
these winding sectors must contribute in order to correspond to the 
standard $O(2)$ model with {\it periodic boundary} conditions. 
We expect, however, that the non-trivial winding sectors are not 
important in the disordered phase. In this case, we would expect 
a correspondence between the standard and the dual formulation both 
with periodic boundary conditions. 

\vskip 0.3cm 
We have verified this by a direct simulation. 
The action density calculated with the $Z$-gauge theory with periodic 
boundary conditions (i.e., zero winding number $n_w=0$) is contrasted to the
result from the $O(2)$ model with periodic boundary conditions in
figure~\ref{fig:5}. As expected, both action densities agree in the 
strong coupling phase for $\beta \ll \beta _c$. For $\beta
\stackrel{<}{_\sim } \beta _c$, the correlation length is sufficiently large
to observe a sizable dependence on the boundary conditions. Large differences
are observed in the  broken phase when windings have a role to play. 

\vskip 0.3cm 
We find that the 3d $O(2)$ (or $xy$) model is dual to a theory of vortices. 
This might explain part of its phenomenological success for solid state 
systems the thermodynamics of which are influenced by vortex 
dynamics (see e.g.~\cite{Kamal1994}).

\section{Quantum field theory - Wigner-Weyl phase }

\subsection{Finite temperatures } 

\begin{figure}
\includegraphics[width=8cm]{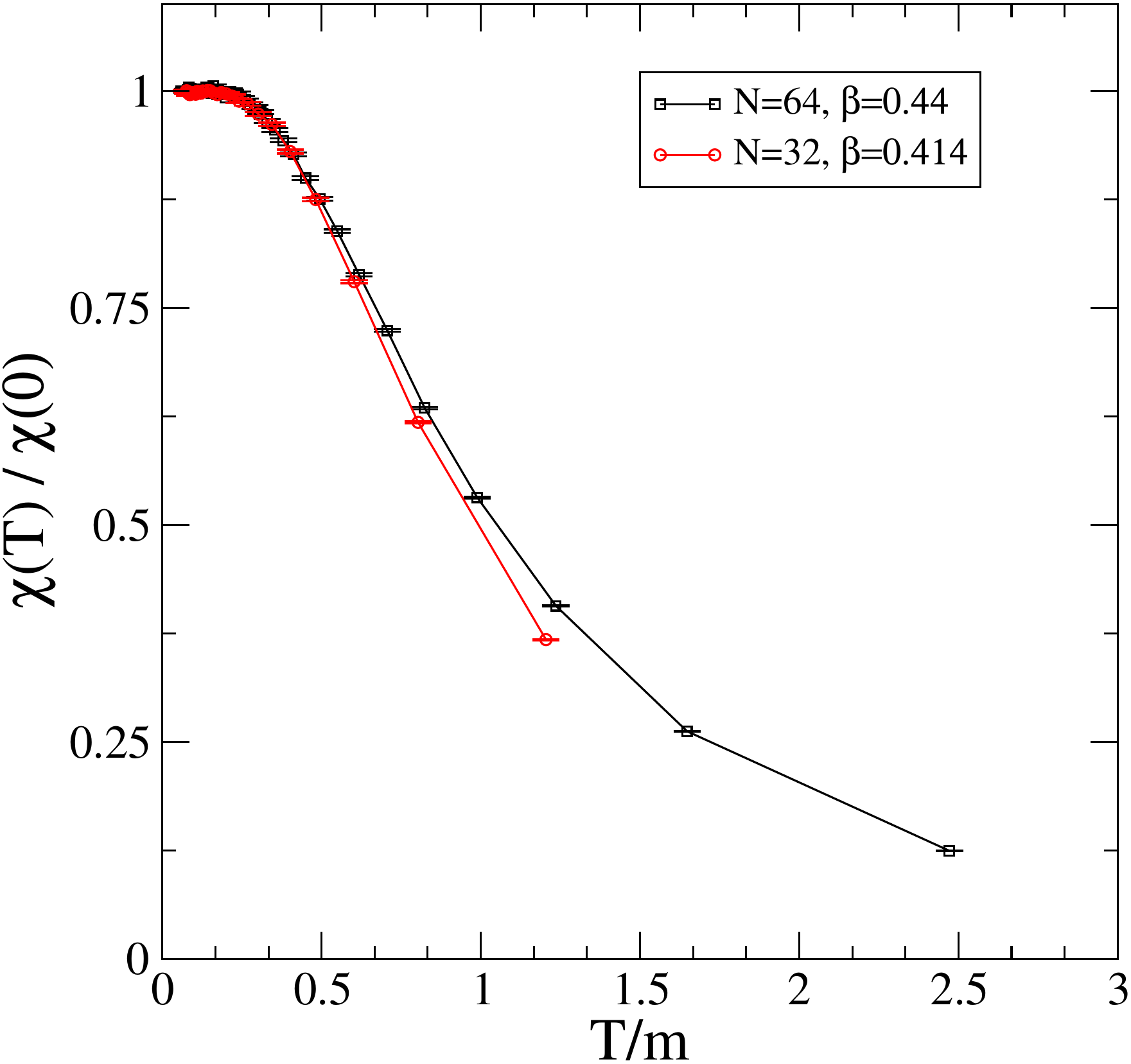}
\caption{\label{fig:7} Magnetic susceptibility $\chi $ in the QFT limit 
of the 3d $O(2)$ model as a function of the temperature $T$ in units of the 
mass gap. }
\end{figure}
Here, we study the emerging quantum field theory in the critical limit 
of the $O(2)$ statistical field theory. We will consider finite 
temperatures, but we will confine ourselves to zero density. 

If we approach the critical coupling $\beta _c$ from below, i.e., 
$\beta \stackrel{<}{_\to} \beta _c$, the $O(2)$ theory remains in the 
Wigner-Weyl phase, while the correlation length $\xi (\beta )$ in units 
of the lattice spacing $a$ diverges. For $\beta $ close to the 
critical coupling, one finds a scaling relation of the
type~\cite{Gottlob:1993jf}:  
\be 
\xi (\beta ) \; = \; \xi_0 \, \cdot \, 
\left(1 - \frac{\beta }{ \beta _c} \right)^{-\nu } \; .  
\label{eq:35}
\en
The correlation length is best calculated using an improved estimator 
and the Wolff algorithm. In fact, it was shown that 
the spin correlation function (\ref{eq:18}) can be estimated 
by~\cite{Hasenbusch:1989kx,Wolff:1989hv}: 
\be 
G(x-y) \; = \; 2 \, \Bigl\langle \frac{N^3}{V_c} \, 
(\vec{r} \cdot \vec{n}_x) \, (\vec{r} \cdot \vec{n}_y) 
\, \theta(C,x) \, \theta(C,y) \, \Bigr\rangle \; ,
\label{eq:36}
\en
where 
$$ 
 \theta(C,x) \; = \; \left\{ \begin{array}{ll}
1 & \hbox{for} \; x \in C \\
0 & \hbox{else.} 
\end{array} \right. 
$$
The scaling parameters have been obtained to good accuracy 
by fitting the scaling relation (\ref{eq:35})
to Monte-Carlo data. Reference~\cite{Gottlob:1993jf} reports 
\bea
\beta _c &=& 0.454157(14), \; \; \; 
\nu \; = \; 0.6711(16), 
\label{eq:37} \\
\xi_0 &=&0.4866(26) \; . 
\nonumber 
\ena 
from an analysis of high precision data. 
For large correlation length, the theory becomes independent from the 
underlying lattice structure and can be effectively described by 
a quantum field theory (QFT). The QFT limit is attained by defining 
a fundamental length scale which plays the role of the only free 
parameter of this theory. Here, we choose the so-called {\it mass gap} $m$ of
the theory which is provided by the inverse correlation length. 
The scaling relation (\ref{eq:35}) then implies 
a dependence of the lattice spacing $a$ on the ``bare coupling'' $\beta $: 
\be 
m\, a(\beta ) = \xi ^{-1} (\beta ) \; . 
\label{eq:38}
\en
The bare coupling $\beta $ is tuned towards the critical coupling. 
This implements the QFT limit by sending the lattice spacing (in physical 
units) to zero. The bare coupling $\beta $ is no longer a free parameter 
of the theory, but its role has been taken over by the dimensionful 
parameter $m$ (dimensional transmutation). In following, we will use 
(\ref{eq:38}) to eliminate the lattice regulator $a(\beta )$ in favour of the
physical mass parameter $m$ and will assume that (\ref{eq:35}) is a good
approximation for the true correlation length $\xi (\beta)$ for the range 
of $\beta $ values used throughout this paper. This assumption is checked by
calculating observables at different renormalisation points $\beta$ and in
physical units. Any violations of the scaling relation would then be detected
by a non-universal behaviour. For the choice of $\beta $ values below, we
found good universal scaling for the observables under discussions. 

\begin{figure}
\includegraphics[width=8cm]{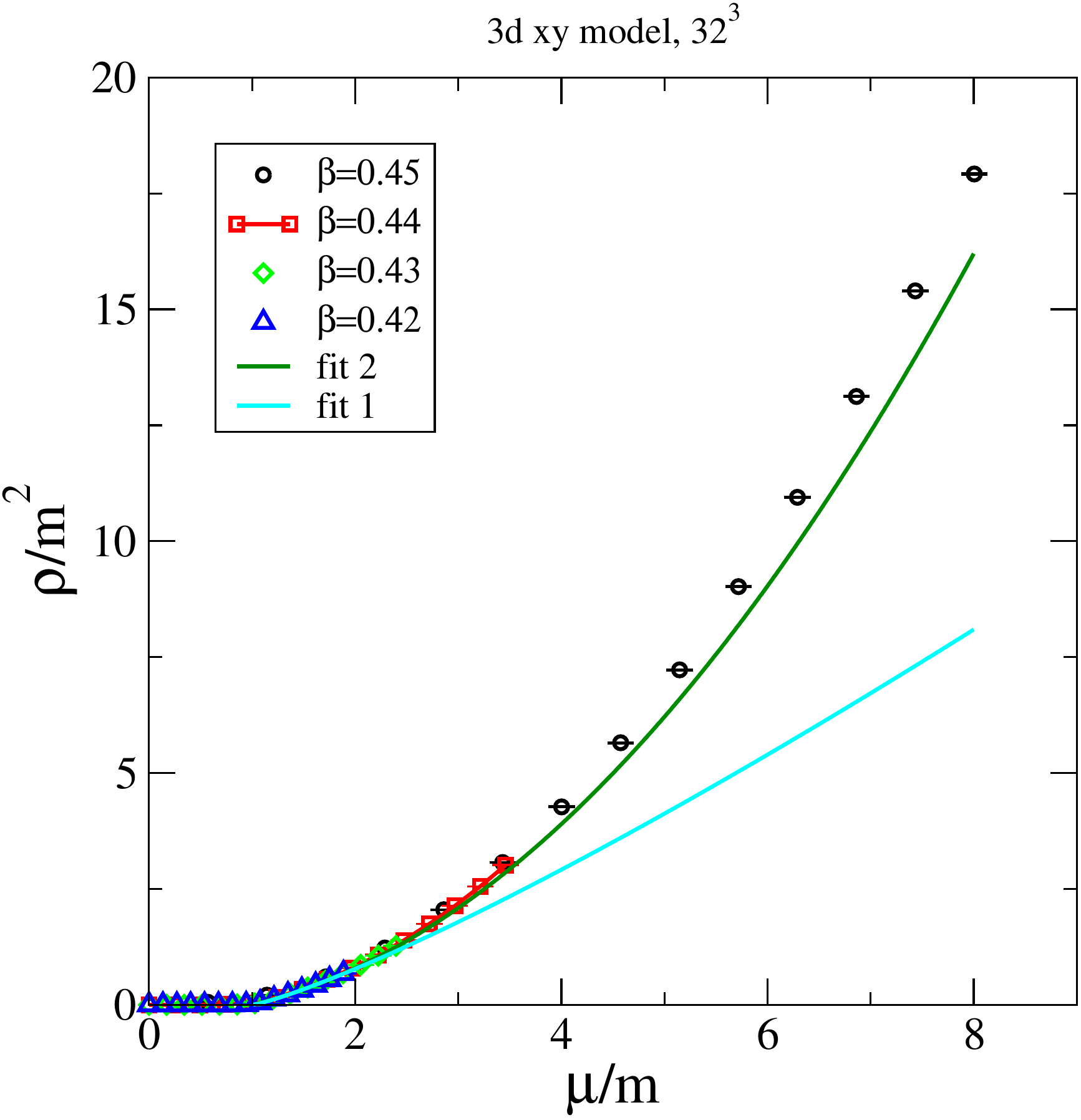}
\caption{\label{fig:6} The charge density $\rho $ as function 
of the chemical potential in units of the mass gap $m$ in the QFT limit
($32^3$ lattice).
}  
\end{figure}
\vskip 0.3cm 
QFT observables are constructed from the ``bare'' observables by 
scaling the quantity of interest in units of the mass gap $m$ 
(to the power of its canonical dimension) and by taking into account 
the wave function renormalisation. Because of its relation to the 
correlation function, the canonical dimension of the 
magnetic susceptibility is two, and we write: 
\be 
Z_\chi (\beta ) \, \chi (\beta)  \; = \; \chi_\mathrm{qft} a^2 \; , 
\label{eq:39}
\en
where $Z_\chi $ is the wave function (or better composite operator)
renormalisation constant. With the help of (\ref{eq:25}), (\ref{eq:35}) and 
(\ref{eq:38}), we find: 
\be 
\chi_\mathrm{qft} / m^2 = Z_\chi (\beta ) \, \frac{  \chi (\beta) }{ 
\xi ^2 (\beta ) } = Z_\chi (\beta ) \, \frac{ \chi _0 }{\xi_0 ^2 } 
\, \Bigl( 1 - \frac{ \beta }{ \beta _c } \Bigr)^{2 \nu - \gamma } \; . 
\label{eq:40}
\en
In a ``minimal subtraction scheme'', we would choose 
\be 
Z_\chi (\beta ) \; = \; \Bigl( 1 - \frac{ \beta }{ \beta _c } \Bigr)^{ - \,
  d_\chi }  \; , 
\label{eq:41}
\en
where the anomalous dimension for this operator $d_\chi $ turns out to be 
quite small: 
\be 
d_\chi \; = \; \gamma - 2 \nu \; \approx \; -0.018(2) \; . 
\label{eq:42}
\en
Needless to say that the actual value of the renormalised quantity 
$\chi_\mathrm{qft}$ must be fixed by a renormalisation condition. 

Finite temperatures, i.e., $T \not= 0$, are implemented by reducing the 
extent of the lattice in the time direction: 
\be 
T \; = \; \frac{1}{ N_t \, a(\beta ) } \hbo 
T/m \; = \; \xi (\beta )/ N_t \; , 
\label{eq:43}
\en
where $N_t$ is the number of lattice points in time direction. 
Once the observable has been renormalised achieving vacuum values which 
enjoy a physical interpretation in the continuum limit $a \to 0$, we can start 
to make predictions at e.g.~finite temperatures. Here, we study the
temperature dependence of the magnetic susceptibility. To this aim, we 
primarily carried out calculations at $\beta = 0.44$ using a $64^2 \times N_t$
lattice. For a study of the lattice spacing (in-)dependence, we also did 
simulations utilising  
a $32^2 \times N_t$ lattice and $\beta = 0.414 $. Note that both simulations 
implement roughly the same spatial volume since 
$$
a(\beta=0.414) \approx 2 \, a(\beta=0.44) \; .
$$
Our findings are summarised in figure~\ref{fig:7}. 
The overall agreement between the two curves is satisfactory bearing in 
mind that the last points at large $T$ correspond to $N_t=2$ for which 
we might expect quite some rotational symmetry breaking effects. 
We observe the magnetic susceptibility stays roughly constant for 
$T \stackrel{<}{_\sim } 0.4 \, m $ while for $T \approx m $ it has fallen 
approximately to half its vacuum value. 

\subsection{Cold, but dense matter } 

\begin{figure}
\includegraphics[width=8cm]{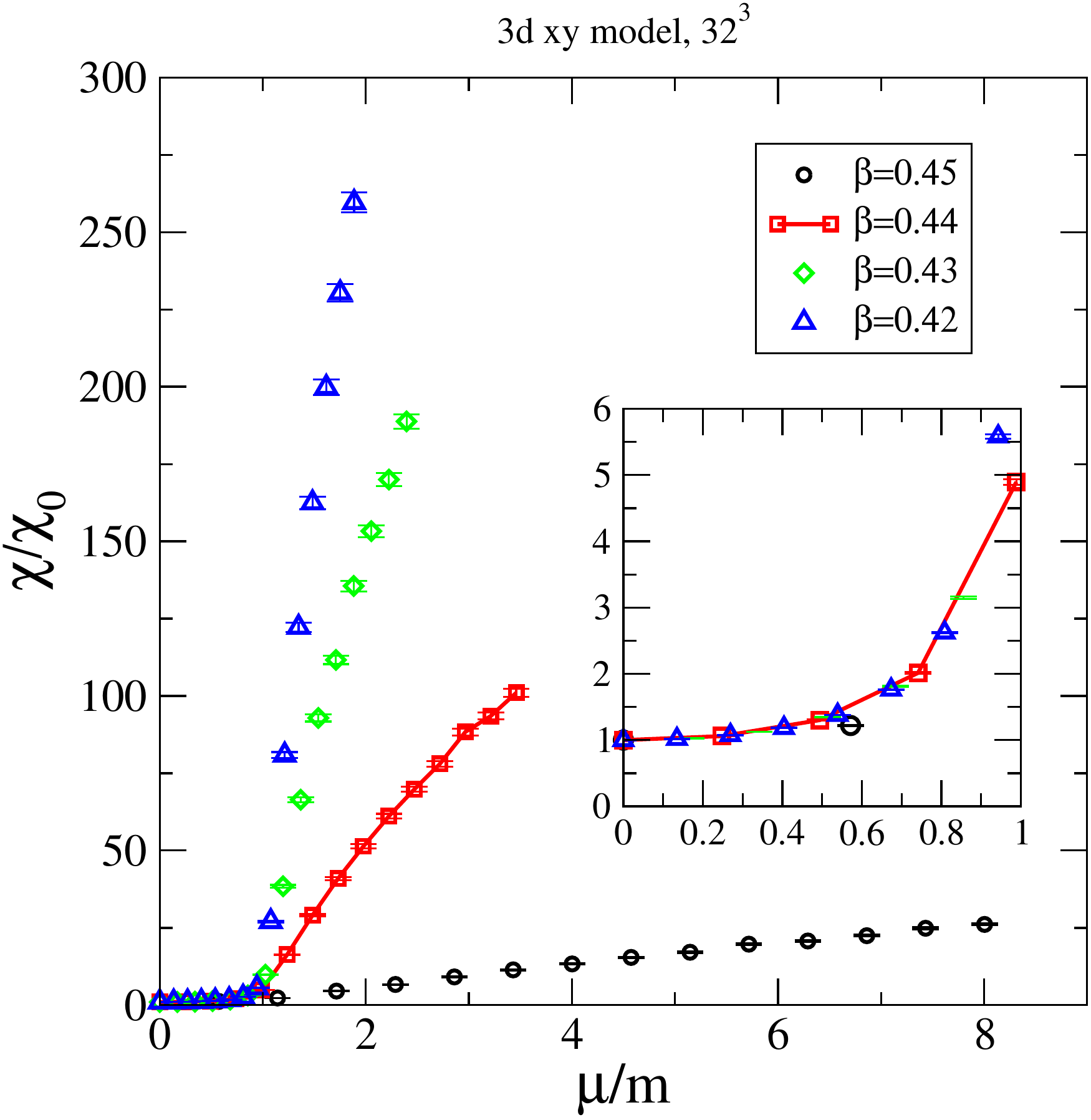}
\caption{\label{fig:8} Magnetic susceptibility $\chi $ in the QFT limit 
of the 3d $O(2)$ model as a function of the chemical potential $\mu$ in 
units of the mass gap.
}  
\end{figure}
\begin{figure}
\includegraphics[width=8cm]{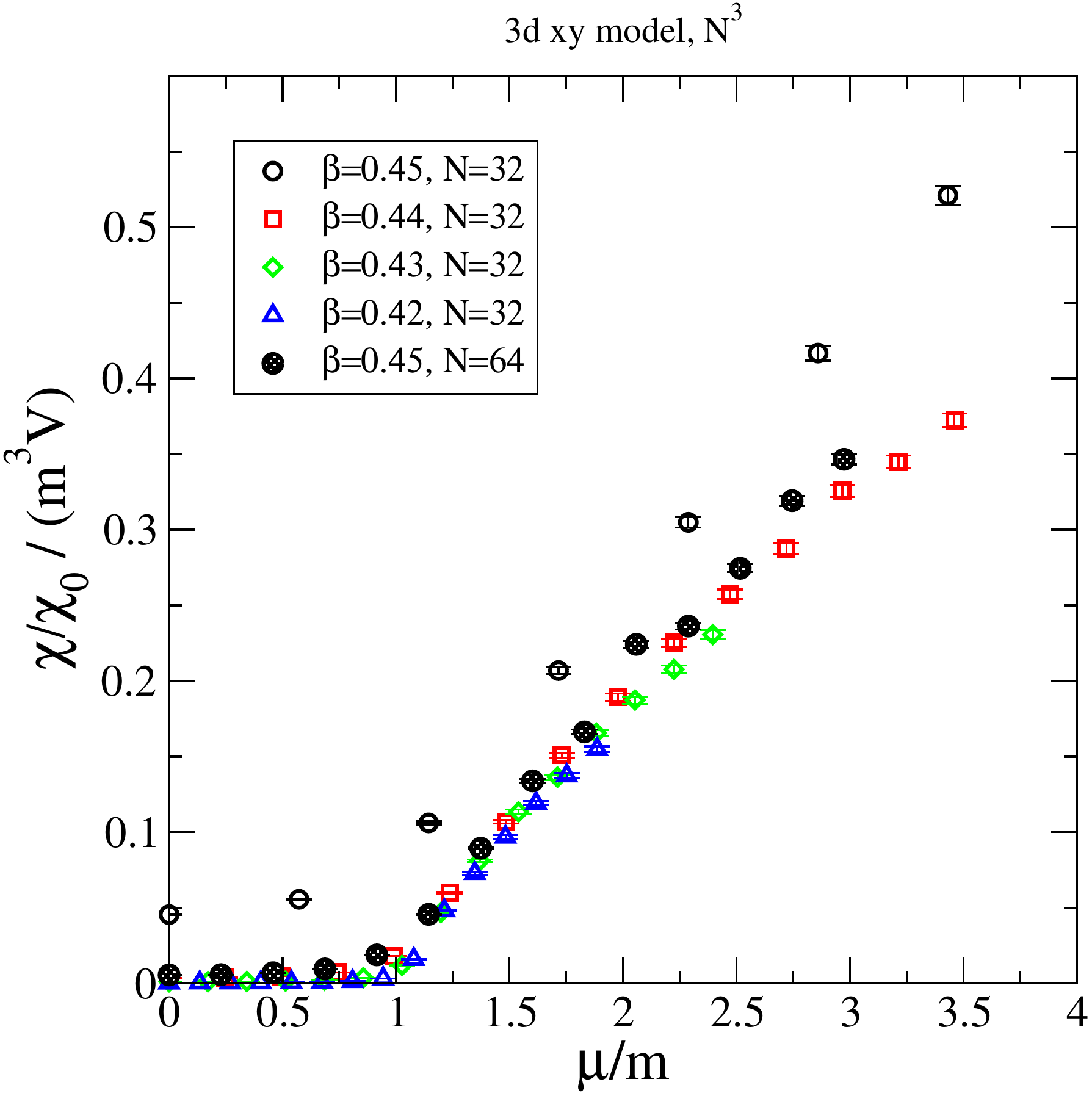}
\caption{\label{fig:9} Magnetic susceptibility $\chi $ {\it per total volume} 
as a function of the chemical potential $\mu$ ($N^3$ lattice). 
}  
\end{figure}
Let us now study finite densities by turning on the chemical potential. 
The physical chemical potential $\mu _\mathrm{phys} $ is related to the bare 
chemical potential $\mu $, the parameter e.g.~featuring in~(\ref{eq:21}), 
by 
\be 
\mu = \mu _\mathrm{phys} \, a \; \; \; \Rightarrow \; \; \; 
\mu _\mathrm{phys} /m \; = \; \xi (\beta ) \, \mu \; . 
\label{eq:52}
\en 
We will do so at zero temperatures, which are adopted by performing simulations 
using a symmetric lattice, $N=N_t$. Again, we study the QFT limit of the 
$O(2)$ theory, but now for non-zero values of the chemical potential. 

The first observable which we will study is the matter density. 
The matter density is easily accessible using the worm algorithm 
(see~(\ref{eq:21})): 
\bea 
\rho &=&
\frac{T}{V_2 } \frac{ \partial \ln Z }{\partial \mu_\mathrm{phys}}  \; = \; 
\frac{1}{V_2 N_t} \frac{ \partial \ln Z }{\partial \mu}  
\nonumber \\ 
&=& \frac{1}{V_2 } \, \Bigl\langle \sum_{x\in \hbox{time slice}} k_\ell 
\Bigr\rangle \; ,  \; \; \; \ell = \langle x,0 \rangle, 
\label{eq:50}
\ena
where $V_2 =  N^2a^2$ is the ``spatial'' volume and the summation extends over
all sites $x$ which are part of a given time slice (see
also~\cite{Banerjee:2010kc}). Due to current conservation, any choice for the
time slice gives the same answer. In physical units, we find 
\be 
\rho /m^2 \; = \; \frac{\xi^2 (\beta ) }{N^2} \, 
\Bigl\langle \sum_{x\in \hbox{time slice}} k_\ell 
\Bigr\rangle \; \; \; \ell = \langle x,0 \rangle . 
\label{eq:51}
\en 
Since the charge is conserved, there is no wave function renormalisation 
of the density. This is confirmed by a direct calculation (see
figure~\ref{fig:6}) using several $\beta $ values ranging from $0.42$ to $0.45$
implying that the corresponding lattice spacing $a$ changes by a factor of
approximately  $4.2$. Yet, the density in physical units nicely falls on top
of a uniform curve. Most importantly, we observe that the density vanishes 
for chemical potentials smaller than the mass gap, i.e., $\mu < m$. 
With other words, the onset chemical potential agrees with the mass gap, and 
the simulation is free of a Silver-Blaze problem~\cite{Cohen:2003kd}. 
Secondly, the density smoothly starts to rise with the chemical potential at
the onset value. This is markedly different from the case of a free Bose gas
for which we would observe that the density diverges logarithmically when
approached from above. This would indicate the well known phenomenon of
Bose-Einstein condensation. However, our data are more compatible with a
quadratic rise of the density with the chemical potential which is
reminiscent of a free 2d Fermi gas  at low temperatures: 
$$ 
\rho _\mathrm{Fermi} \propto (\mu -m)^2 \; . 
$$
Although the standard $O(2)$ model is formulated in terms of bosonic 
degrees of freedom, it is not ruled out that the model near the onset
transition  is well described 
by an effective fermion theory. To trace this out further, we have fitted the 
numerical data in figure~\ref{fig:6} to a simple scaling law: 
\be 
\rho /m^2 \; = \; a_0 \, (\mu/m -1)^{a_1} \; . 
\label{eq:60}
\en 
We find, however, that the data are not well represented by this ansatz. 
Performing independent fits to each set for a given $\beta$, the results 
are collated in the first two rows of table~\ref{tab:1}: 

\begin{table}[tbh]
\begin{tabular}{r|rrrr}
 & $\beta = 0.42 $ & $\beta = 0.43 $ & $\beta = 0.44 $ & $\beta = 0.45 $ \\
 \hline 
 $a_0$ & $0.77(1)$ & $0.82(1)$ & $0.827(5)$ & $0.70(2)$ \\
 $a_1$ & $1.2(1)$  & $1.27(2)$ & $1.42(2)$  & $1.66(2)$ \\
 $b_0$ & $0.54(2)$ & $0.55(2)$ & $0.55(2)$ & $0.58(1)$ \\
 $b_1$ & $0.25(3)$  & $0.26(2)$ & $0.27(1)$  & $0.281(6)$ \\
\end{tabular}
\caption{\label{tab:1} Fitting the density to the power law ansatz 
(\ref{eq:60}) and to (\ref{eq:61}). }
\end{table}

We point out that  the data are quite well fitted by the ansatz 
\be 
\rho /m^2 \; = \; b_0 \, \mu/m \; + \; b_1 \, (\mu/m)^2 \; . 
\label{eq:61}
\en 
The coefficients $b_0$, $b_1$ are also listed in table~\ref{tab:1}. 
Figure~\ref{fig:6} shows the curves from the fit according to
(\ref{eq:60}), called  fit 1, and according to (\ref{eq:61}) called fit 2. 
The fitting curves have been generated using the fit parameters obtained
for $\beta =0.42$. 

\vskip 0.3cm

\vskip 0.3cm 
\begin{figure*}
\includegraphics[width=8cm]{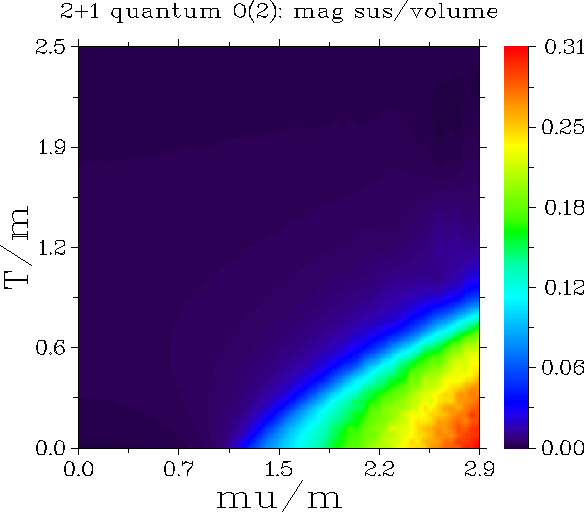} \hspace{0.3cm}
\includegraphics[width=7.7cm]{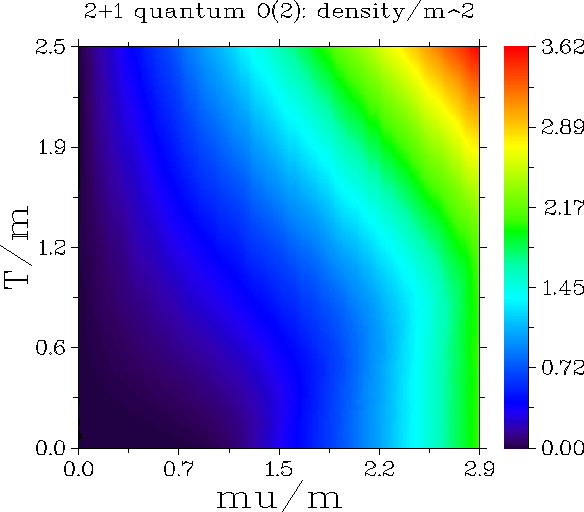}
\caption{\label{fig:10} Phase diagram: Magnetic susceptibility $\chi $ 
{\it per total volume} (left) and density (right). 
}  
\end{figure*}

It is interesting to note that a free Fermi gas at small, but non-zero 
temperature would suggest the $\mu $-dependence shown in
(\ref{eq:61}). In this case, the fitting coefficient $b_0$ should be 
proportional to the temperature. In order to explore the potential 
correspondence between the $O(2)$ quantum model and a Fermi gas, 
further simulations for different lattice sizes and aspect ratios are 
necessary. This is left to future studies. We also point out that 
in the case of a free gas of hardsphere
bosons~\cite{PhysRevB.65.104519} one expects a linear rise of the
density with the chemical potential $\mu \stackrel{>}{_\sim } \mu
_c$. Note, however, that the coefficient $b_1$ should be negative in
the latter case.

\vskip 0.3cm 
Let us finally explore the magnetic susceptibility $\chi $ below and above the
onset  value of the chemical potential. Figure~\ref{fig:8} shows $\chi $ 
normalised to its zero temperature value $\chi _0$ in order to remove 
the renormalisation constant. The inlay of this figure shows that 
$\chi / \chi_0$ is indeed independent of the renormalisation point 
specified by $\beta $. Although the density remains zero for $\mu < m$, 
the chemical potential impacts on the properties of the theory as indicated by 
the increase of $\chi / \chi_0$ with $\mu$. For $\mu >m$, the data for 
$\chi / \chi_0$ largely depend on the value $\beta $. Note that the
simulations (except for $\beta =0.45$) are carried out for a fixed number of
lattice points. Here, a change of $\beta $ also implies a change of the
lattice volume.  
This is indeed the cause for the apparent scaling violations:
Figure~\ref{fig:9} shows the magnetic susceptibility per volume,
i.e., $\chi / \chi_0 / (m^3 V)$. The volume $V$ is the physical volume and 
the mass gap is introduced to render this factor dimensionless. The 
factor $\chi _0$ is again necessary to remove the renormalisation constant. 
We observe good scaling for the magnetic susceptibility in units of 
the (physical) volume for a wide range of $\beta $ values. An exemption 
is the result for $\beta = 0.45$. For this $\beta $, the correlation length
equals roughly a third of the system size and finite volume effects come into
play. At $\beta =0.45$, we also calculated  $\chi / \chi_0 / (m^3 V)$ for a 
$N^3=64^3$ lattice. A good agreement with the results from smaller $\beta $ and
for a $N^3=32^3$ lattice is recovered. We interpret the physics behind 
these findings as follows: For $\mu>m$, matter starts populating the ground
state. This alters the interaction between the spins such that the U(1) global
symmetry of the model breaks spontaneously leading to a superfluid phase.

\vskip 0.3cm 
We are finally in the position to calculate the phase diagram of the $2+1$
dimensional quantum $O(2)$-theory in the $\mu /m$ and $T/m$ plane. 
We have carried simulations with $\beta =0.44$ using a lattice size of 
$64 \times N_t$, where $N_t=2 \ldots 32$. We performed simulations for 
$\mu a = 0.01 \ldots 0.6$ in steps of $0.01$. Naturally, the temperature steps 
from (\ref{eq:43}) are not equally spaced. We used a cubic spline
interpolation to generate a regularly spaced $60 \times 60$ data set
before plotting.  

\vskip 0.3cm 
To detect the superfluid phase in the phase diagram, we plotted 
the renormalised magnetic susceptibility in units of the physical volume,
i.e., $\chi / \chi_0 / (m^3 V)$ where $V$ is the (physical) volume of the 
3-torus, i.e., $V = L^2/T$. The result is shown in figure~\ref{fig:10}, left
panel.  
As expected, we find the superfluid phase in the cold, but dense region. 
Increasing temperature at a fixed chemical potential $\mu > m$, 
rapidly dissolves the superfluid phase. Also shown is the density in the 
$\mu$-$T$-plane. At low temperatures, we observe the onset at 
$\mu \ge m$. For $\mu \approx m$, we observe an increase of the density with 
temperature, which is due to thermal excitations with more 
particles overcoming the mass gap than anti-particles. 

\section{Conclusions} 

Using existing simulations techniques such as the Wolff cluster 
algorithm and the worm algorithm, we have thoroughly analysed the 
3d $O(2)$ model in the continuum limit $\beta \to \beta _c$. 
In this limit, the model represents the quantum $O(2)$ model in 
$2+1$ dimensions. The mass gap $m$ takes over the role of the free (scale) 
parameter, while the lattice spacing $a$ is fine-tuned 
by the $\beta $ dependent correlation length, $m a(\beta) = 1/\xi (\beta)$. 

\vskip 0.3cm 
We have analysed the zero density limit by calculating the 
(renormalised) magnetic susceptibility as a function of the temperature. 
As usual, the temperature depends on the extent $N_t$ of the torus in time 
direction, i.e., $T/m = \xi (\beta)/N_t$. We find good scaling 
over more than a factor of $4$ in the lattice spacing. 
The magnetic susceptibility is rather independent of the temperature 
for $T<m/2$ while a rather steep descent is observed for 
$T \stackrel{>}{_\sim} m $. 

\vskip 0.3cm 
Charge is defined by means of the Noether current of the global $U(1)$ 
symmetry of the theory, and finite densities are introduced using 
the chemical potential $\mu $. In a first step, we analysed the $\mu $ 
dependence of the (renormalised) magnetic susceptibility and the density 
at ``zero'' temperature (isotropic lattice). Simulations are carried out 
using the worm algorithm. The onset chemical potential $\mu _c$ is found to
agree  with the mass gap $m$ of the theory. At $\mu \stackrel{>}{_\sim} 
\mu_c$, the density smoothly rises with increasing $\mu $ ruling out 
Bose-Einstein condensation. In fact, the $\mu $ dependence of the density is
more in line with a free Fermi gas (at small temperatures). More 
investigations are needed to understand the theory close to onset in terms 
of an effective theory. This is left to future work. 

\vskip 0.3cm 
For $\mu \ge \mu _c$, the magnetic susceptibility scales with the volume 
and signals the spontaneous breakdown of the $U(1)$ symmetry and therefore 
superfluidity. We finally present the results for the full $\mu $-$T$ phase
diagram. In the cold, but dense regime we find a superfluid phase, which 
is rapidly resolved by increasing the temperature.

\begin{acknowledgments}
We would like to thank Shailesh Chandrasekharan,  Christof Gattringer and 
Kenji Fukushima for helpful discussions. This research is supported by STFC
under the DiRAC framework. We are grateful for support from the HPCC Plymouth,
where the numerical computations have been carried out. 
\end{acknowledgments}

\bibliography{xy_stan}

\end{document}